\documentclass[conference]{IEEEtran}
\usepackage{graphicx}

\ifCLASSINFOpdf
\else
\fi
\usepackage{flushend}

\usepackage{listings}

\begin{document}
%
\title{Inherent Diversity in Replicated Architectures}

\author{\IEEEauthorblockN{Peter Okech}
\IEEEauthorblockA{Department of Physics\\
University of Nairobi\\
Nairobi, Kenya\\
Email: pokech@uonbi.ac.ke}
\and
\IEEEauthorblockN{Nicholas Mc Guire}
\IEEEauthorblockA{OpenTech Research EDV\\
Bullendorf, Austria\\
Email: der.herr@hofr.at}
\and
\IEEEauthorblockN{William Okelo-Odongo}
\IEEEauthorblockA{School of Computing \& Informatics\\
University of Nairobi\\
Nairobi, Kenya\\
Email: wokelo@uonbi.ac.ke}}

\maketitle

\begin{abstract}
In this paper, we report our ongoing investigations of the inherent non-determinism in contemporary execution environments that can potentially lead to divergence in state of a multi-channel hardware/software system. Our approach involved setting up of experiments to study execution path variability of a simple program by tracing its execution at the kernel level. In the first of the two experiments, we analyzed the execution path
by repeated execution of the program. In the second, we executed in parallel two instances of the same program, each pinned to a separate processor core. Our results show that for a program executing in a contemporary hardware/software platform, there is sufﬁcient path non-determinism in kernel space that can potentially lead to diversity in replicated architectures. We believe the execution non-determinism can impact the activation of residual systematic faults in software. If this is true, then the inherent diversity can be used together with architectural means to protect safety related systems against residual systematic faults in the operating systems.
\end{abstract}


\section{Introduction}\label{sec_intro}

Modern computing platforms that provide an execution environment for user applications, comprising of the hardware and the operating system, are becoming increasingly complex. Contemporary processor possess features such as pipelines, branch prediction, several levels of caches, out of
order execution, frequency scaling, and multiple cores. These features, while making the processor architecture complex, also introduces a certain level of perceived indeterminism during program execution. The operating system, the system software that manages the hardware and provide services to user applications, are also growing in complexity, as evident from growth in size of the GNU/Linux operating systems from 2 MLOC in 2001 to more than 9 MLOC in 2011~\cite{palix2014faults}.

An increasing number of applications are now being classified as safety-related. Consider, for example, a software based turn indicator controller, an application with simple logic that performs some safety functions. The developer of the safety application has several options available: She could 1) make use of specialized hardware/micro-controller, and develop the complete software stack consisting of the application and the execution runtime using the prescribed process of a relevant safety standard or 2) choose to make use of a platform based on a modern processor architecture and an off-the-shelf operating system such as GNU/Linux.

Suppose the developer has chosen the second option, and the following two assumptions hold: First, the use of contemporary hardware is acceptable from the safety certiﬁcation perspective and secondly, the safety application is considered correct and error free by virtue of it having been developed and verified in accordance with the procedures and processes of a given safety standard.

The primary challenge that the developer would be faced with is that of ensuring that even with the presence of residual faults in a complex operating system not developed within a safety context, the safety application will not have dangerous failures. We suggested a fault detection mechanism based on redundancy and diversity that harnesses the nondeterminism in the behavior of a complex system.

Our work is based on the premise that rather than fight complexity present in the execution environment, we can exploit it to achieve the goal of ensuring that residual faults in the software platform do not lead to the failure of user applications. In this respect, we propose to take advantage of the inherent non-deterministic execution in complex execution platforms to generate diversity at runtime in replicated homogeneous architectures. This inherent diversity approach, together with architectural means, could be used for fault detection in order to provide protection for safety-related systems executing in contemporary hardware/software platform.

Our work is guided by the safety standard IEC 61508~\cite{iec201061508Ed2} that, among other prescriptions, suggests the use of architectural protection to provide coverage for faults during operations of safety-related systems.

The design for fault tolerance has two main aspects: building of a redundancy structure to cope with faults and the verification of the effectiveness of the fault tolerance mechanism~\cite{strigini2005fault}. At this point in our study, we focus on the first aspect, that is the definition of a fault detection mechanism for residual design faults in the operating system in the context of safety. The work reported in this paper builds upon our previous work~\cite{okech2013investigating, okech2014utilizing} on individual systems calls, by extending the analysis to a set of system calls corresponding to some user application task. We first present the results of analyzing the level of variability of kernel execution paths and then give our interpretation of the results. Specifically, we:
\begin{itemize}
	\item Show that a code fragment executed repeatedly displays variability of its execution path in kernel space.
	\item Demonstrate that replicated threads executing on separate cores of a dual core processor follow divergent paths through the kernel space.
	\item Argue that the diversity in replicated architecture due to nondeterminism in the execution platform can be used, for example in 2-out-of-2 conﬁguration, for fault detection.
\end{itemize}

The rest of the paper is organized as follows: The next section, Section~\ref{sec_back} provides a some background information related to our work. Section~\ref{sec_method} describes the methods and procedures used in the experiments. We then follow this by describing the experimental study (Section~\ref{sec_study}) and the results
obtained (Section~\ref{sec_results}). In Section~\ref{sec_discuss}, we give our interpretation of the results. Related work is discussed in Section~\ref{sec_related}. We then offer a conclusion in Section~\ref{sec_concl}.

\section{Background}\label{sec_back}

A typical user application, such as a safety-related program, executes in an environment comprising of the operating system and the hardware. Contemporary computing platforms are becoming increasingly complex, making it difﬁcult to assure deterministic properties of an application during its execution.

The computing execution platform operate through the interaction of its software and hardware components. The interaction leads to coupling of the elements of the systems, resulting into global properties that cannot be attributed to any of the components in isolation. This leads to some level
of unpredictability in the internal behavior of an application executing in such complex platforms with respect to the paths and timing of system services. In the absence of errors, this unpredictability does not impact the correctness of the system function.

Faults activated at runtime may lead to erroneous system states or errors, which might eventually manifest as failure at the system level. The main approach to dealing with failures during systems operation is the use of redundancy to achieve fault tolerance. In addition, for failures that might lead to catastrophic consequences, the general approach is to use fail-safety to minimize the impact of failure.

The predominant faults in software systems can be attributed to design faults, arising from either specification, design or implementation. Diversity in design was suggested in the 70's~\cite{randell1975system, avizienis1977implementation}, as a solution to common mode failures that might arise from implementing redundancy by naive replication of software.

However, the high cost of generating diversity at design time has made it a preserve of only very high integrity systems. Automated diversity techniques that removes the need for manually generating diverse versions of software, for example through the use of compilers, has been suggested as a means of cost-effectively generating diversity. An in depth survey of these and related techniques can be found in~\cite{baudry2014multiple}.

Another possibility of generating diversity is to take advantage of non-determinism in the complex execution environment. This is what we refer to as inherent runtime diversity, which provide a promising means to protect systems against residual design faults in complex software systems. 

Speciﬁcally, we focus on the non-determinism in the execution path of an application in kernel space. We seek to (1) determine if there is indeed non-determinism in the execution environment, and (2) if it is sufficient to manifest as diversity in a general N-out-of-M (NooM) architecture with a voting mechanism.

\section{Approach}\label{sec_method}

The first step in our study was to develop a program that invokes the services of the operating system. In our test scenario, the main task of the program was to read a value from a text ﬁle, and then write back the value read to the same location in the file. These basic steps were performed within a repetition loop of 1000 iterations. The pseudocode of Listing~\ref{lst_pseudo} illustrates the steps of the task of the test program. Deliberately, we do not manipulate the value read from the ﬁle before writing it back to ensure that the same user level input across all the iterations of the program. We chose to implement the main task as a Posix thread in a c-language program.

\begin{lstlisting}[caption = Pseudo Code of the Test Program, basicstyle=\footnotesize, label=lst_pseudo]

	...
	fd = open ( ... )
	while ( iter > 0 ) {
		read ( ... )
		lseek ( ... )
		write ( ... )
		fsync ( ... )
		lseek ( ... )
		sleep ( ... )
		iter = iter - 1	
	}
	...
\end{lstlisting}

To enable us collect data on the execution path behavior of the application when executing in kernel context, we conﬁgured the kernel to support FTrace~\cite{rostedt2009world}, a tracing utility that is built into the Linux kernel. At each execution of the program, the tracer recorded the call graphs, i.e. the set of kernel routines that were executed when the application invoked a system call.

We designed two experiments in order to study the variability of the execution path of the program in kernel space. In our work, we define the path of a program as the collection of kernel functions that are called by the set of system calls invoked from the user space by the program. The metric that we use to identify a path is the count of the number of kernel functions (we also regard the system call as a function) which we refer to as the length of the path.

There are several factors that might influence the variability of the execution properties, such as execution time and kernel execution path, of a program. One of these is the placement of a program's thread of execution on one of the available processing units. Studies that investigate the effect of thread placement, referred to as processor affinity strategies such as~\cite{mazouz2010measuring}, \cite{nogueira2014experimental} tend to focus on how to ensure determinism in a programs execution time. With our interest being on setting up a logical 2-channel software/hardware system on a mulicore machine, we first had to investigate the effect of thread pinning on the kernel execution path variability.

In experiment \#1, the study involved comparing the paths of repeated execution of the program. In addition, we considered two thread placement options: 1) No affinity, whereby the thread is dynamically allocated to an available processor core by the operating system scheduler and 2) Processor affinity, whereby the thread is pinned to a particular processor core by the programmer. This strategy results in the creation of two experimental groups, the ``free'' group and ``pinned'' groups for option 1 and 2 respectively. Other than observing the variability , we also performed statistical analysis to determine whether the paths from the sample execution runs from the ``free'' and ``pinned'' are drawn from the same distribution - i.e. can be considered as practically equivalent.

In experiment \#2, our primary aim was to develop an architectural configuration with two program replicas executing in parallel and observe the differences, if any of the kernel execution paths of the replicas. Our approach was to have two copies of the same thread pinned to different processor cores executing simultaneously while recording a trace of their executions.

We used repeated experimental approach for both experiments \#1 and \#2. Specifically, we performed each experiment 5 times resulting in experimental runs which we labeled as
runs A, B, C, D and E.

\section{Experimental Study}\label{sec_study}

\subsection{Experimental Environment}\label{sec_env}

All the programs were executed in an Intel Core 2 Duo machine running Debian GNU/Linux, kernel version 3.16.1. The experimental set up is shown in Table~\ref{tab_spec}. The kernel was configured for tracing using the Ftrace tool.

\begin{table}[!t]
	\caption{Experimental Machine Specification}
	\label{tab_spec}
	\centering
	\begin{tabular}{|l|c|}
		\hline
		\multicolumn{2}{|c|}{HP Compaq}\\
		\hline
		\hline
		Processor & Intel(R) Core 2 Duo E8400\\
		\hline
		No of Cores & 2\\
		\hline
		CPU Speed & 3.00GHz\\
		\hline
		RAM & 2 GB\\
		\hline
		OS & Debian 7.0, Linux Kernel 3.16.1\\
		\hline
	\end{tabular}
\end{table}

The test program was executed in an idle system with no load other than background daemons and the operating system. For the repeated experiments, we rebooted the machine before performing a new experimental run to ensure that each experiment was conducted in a fresh execution environment.

\subsection{Experimental Description}\label{sec_desc}

For experiment \#1, the program described by the pseudocode of Listing~\ref{lst_pseudo} was compiled using the gcc compiler defaults and the NPTL thread library. We then executed the program, via trace-cmd~\cite{rostedt2010trace}, a front-end the FTrace tool to record the traces of the kernel functions called by the executing program. We then analyzed the call graphs of the individual systems calls within the program task, and then the task statements as a single unit.

To study the effect of pinning the thread to a processor core, we modified the program by specifying the thread affinity and repeated the experiment.

In experiment \#2, we first modified the program to create two instance of the thread that performs the program task. These two threads were pinned to one of the two available processor cores (cores [000] and [001] respectively) in the experimental machine. We then executed the program and recorded the kernel execution traces. For this experiment, we did not perform analysis for individual system calls.

\section{Results}\label{sec_results}

As mentioned in section~\ref{sec_method} we used repeated experiment methodology in our work. The values of the measurements obtained displayed variability, as expected. For brevity, we present only one set of results from the experimental runs unless there is need for comparing the measurement values of all the five runs. To avoid bias we have chosen to present, from the possible five runs, the results of the third experimental run (experimental run C) across the two experiments.

\subsection{Experiment 1}\label{sec_exp1}

The individual system calls that were invoked from the user space by the thread's loop are in the system call set \{read, write, lseek, fsync, nanosleep\} and signals \{rt\_sigprocmask, rt\_sigaction\}. We extracted the summary statistics of the ``free'' experimental runs. The values of the minimum, maximum, average, median, mode and the standard deviation of the length of execution paths for each of the system calls and signals for experimental run C are given in Table~\ref{tab_summary}.

\begin{table}[!t]
	\caption{Summary Statistics of Path Lengths for a Sample Run}
	\label{tab_summary}
	\centering
	\resizebox{\columnwidth}{!} {
	\begin{tabular}{|l|r|r|r|r|r|r|}
		\hline
		 & \multicolumn{6}{|c|}{Length of Paths}\\
		\hline
		\hline
		 & Shortest & Longest & Ave. & Median & Mode & STDV\\
		\hline
		read & 91 & 888 & 96.8 & 490 & 91 & 32.3\\
		\hline
		lseek & 13 & 124 & 14.4 & 28 & 13 & 7.5\\
		\hline
		write & 149 & 395 & 159.1 & 272 & 149 & 38.5\\
		\hline
		fsync & 435 & 1069 & 505.2 & 649 & 484 & 73.0\\
		\hline
		lseek & 13 & 250 & 13.3 & 132 & 13 & 7.6\\
		\hline
		rt\_sigprocmask & 7 & 175 & 7.2 & 7 & 7 & 5.4\\
		\hline
		rt\_sigaction & 3 & 23 & 3.1 & 3 & 3 & 1.2\\
		\hline
		rt\_sigprocmask & 7 & 103 & 7.3 & 17 & 7 & 3.4\\
		\hline
		nanosleep & 71 & 187 & 73.6 & 94 & 71 & 12.7\\
		\hline	
	\end{tabular}
}
\end{table}

Examining the paths of these individual system call revealed the characteristics of different categories of systems calls. Some of the simple system calls and signals (for example lseek, nanosleep, rt\_sigprocmask, and rt\_sigaction), which call a small number of kernel routines, display stable/predictable behavior with an occasional `spike' in the number of functions called. The most common invocations of these systems calls involved execution of the smallest set kernel routines, as represented by the matching values of shortest and mode in Table II. The median and average values close to the shortest path lengths, and the small values of the standard deviation for these system calls indicate that it is likely that when these system calls are invoked, the execution instance would take a `typical' path.

There are greater variability in the length of the execution paths for the non-trivial system calls, such are the read(), write() and fsync(). These system calls show a large standard deviation (last column of Table~\ref{tab_summary}) compared to the other system calls. Of interest to us was the occurrence frequencies of each path of unique length for these system calls from the 1,000 execution instances. For example, we identified 38 unique paths for the write() system call, with the highest occurrence frequency of 830 for the path length of 149. 35 of the 38 paths had frequencies of 7 or less. Similarly, we looked at the path variability of the fsync() system call. From the 154 unique paths, 129 had occurrence frequencies of 10 paths or less. The highest frequency was 92 execution instances for path length of 484.

In addition to analyzing the length of the execution paths of individual system calls, we also considered the statements in the thread’s task loop as single unit. The task performed by such a unit is what is typically used to implement the control and/or safety functions in real world applications. Furthermore, for multi-channel systems, it is the value(s) that are produced at the end of such a control loop, that are usually used for adjudication.

\begin{figure}[!t]
\centering
\includegraphics[width=0.48\textwidth]{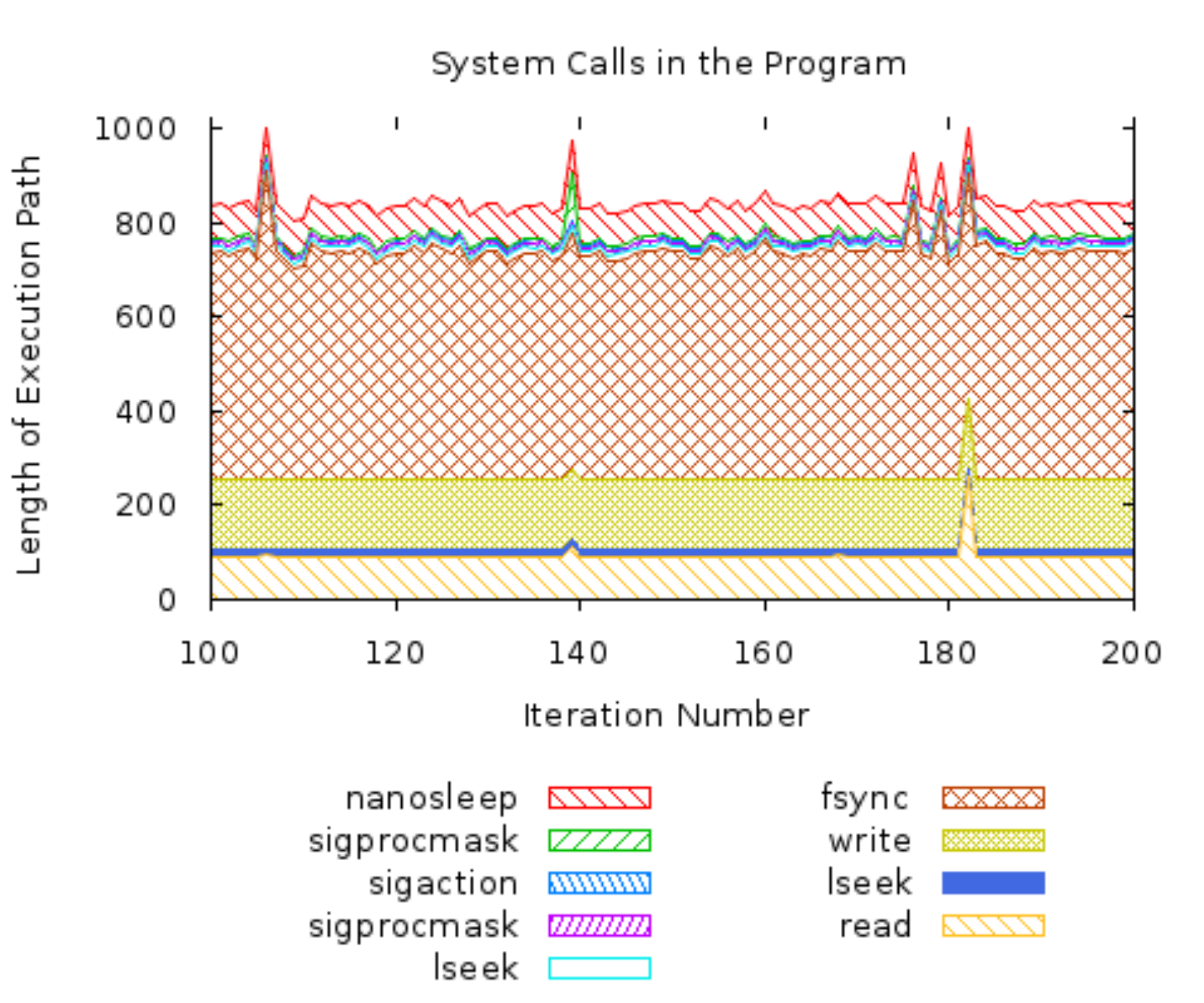}
\caption{Variation in the Length of Execution Path per System Call (snapshot)}
\label{fig_snap}
\end{figure}

We therefore analyzed the data to observe the variability of the task loop statements' execution paths. Fig.~\ref{fig_snap} is a snapshot of the execution path of the task as a composition of the loop statements from the individual system call, showing the contribution of each of the system calls to the overall path variability.

\begin{figure}[!t]
\centering
\includegraphics[width=0.48\textwidth]{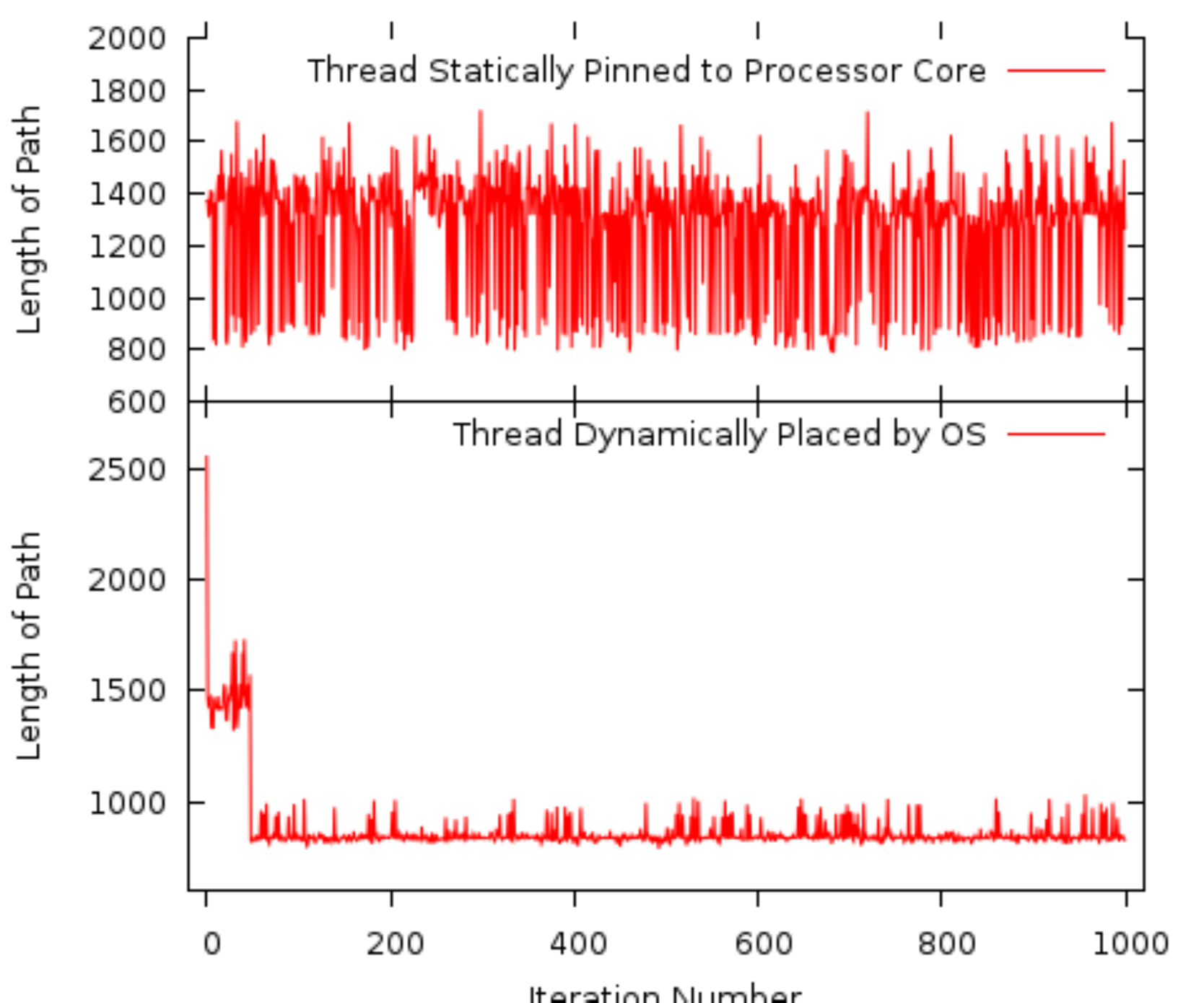}
\caption{Variations in the Length of Execution Paths for the ``free'' (lower figure) and ``pinned'' (upper figure) Experimental Groups}
\label{fig_fp}
\end{figure}

Our next task involved comparing the variability of the kernel execution paths of the thread from the ``free'' and ``pinned'' experimental runs. For this purpose, we plotted the length of the paths for each iteration from the two experimental runs. This is given in Fig.~\ref{fig_fp}. The path lengths of the execution runs from the ``pinned'' groups showed more variability and spread (larger difference between the upper quartile and the lower quartile values), but less extreme values compared to those from the ``free'' group.

\begin{figure}[!t]
\centering
\includegraphics[width=0.48\textwidth]{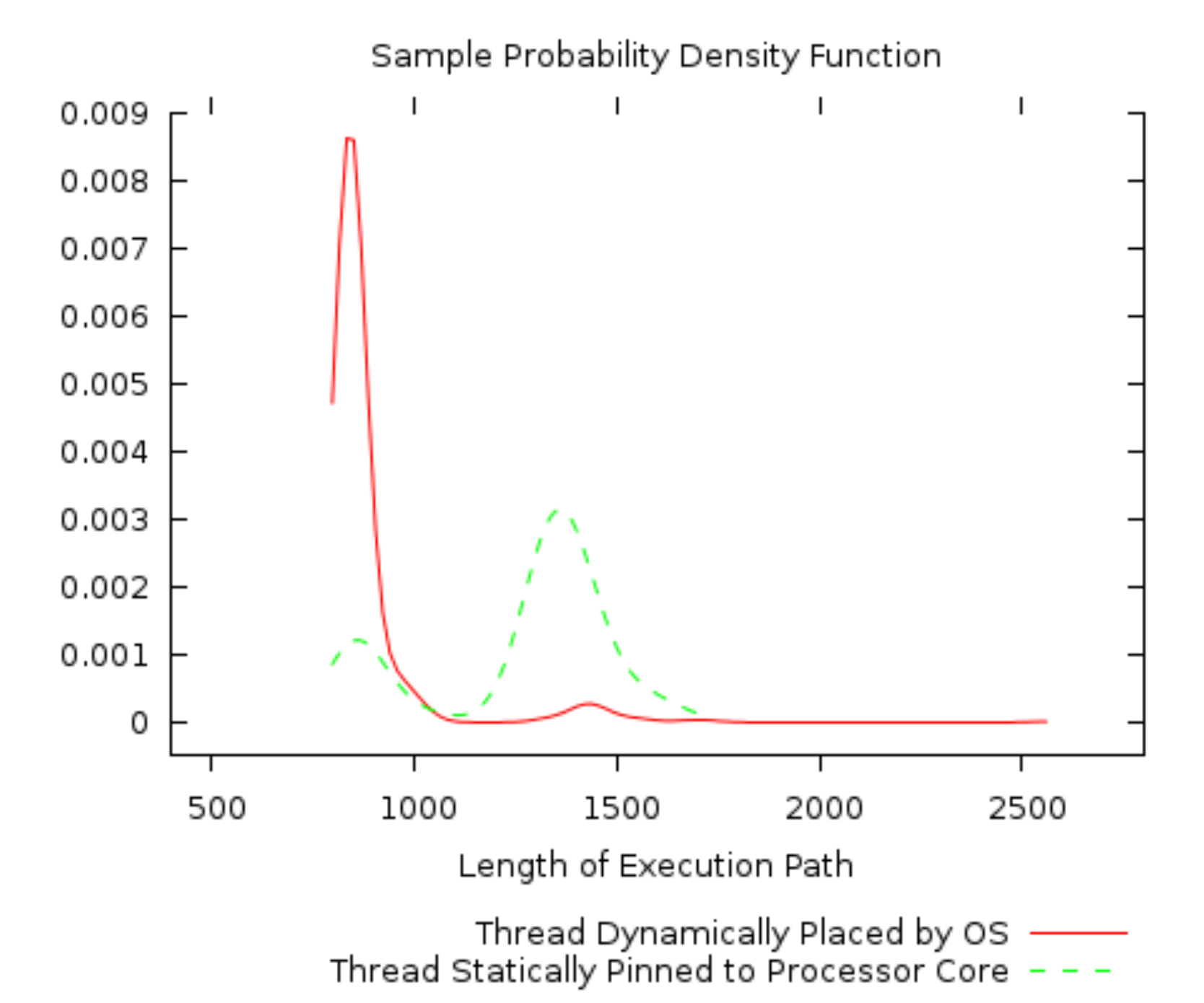}
\caption{Probabiity Distribution for the Sample}
\label{fig_kdf}
\end{figure}

Further, to get a better insight of the differences between the two groups of experimental runs we generated a kernel density distribution of the length of execution paths for the respective experimental runs. Fig.~\ref{fig_kdf} shows the graph of the sample probability distribution for the ``free'' and ``pinned'' experimental groups for run C. The probabilty distributions displayed the same shape, i.e. bi-modal, with varying parameters.

We performed statistical test, using the two sided test of equivalence, to check whether at 0.05 confidence level, we can state that two samples, one from ``free'' group and one from the ``pinned'' group, come from the same population. The results for the experimental run C of the ``free'' group compared with the corresponding experimental run (run C) from the ``pinned'' group is shown in Table~\ref{tab_equiv}.

\begin{table}[!t]
	\caption{Test of Equivalence for Free and Pinned Groups - Run C at 95\% CI, Similarity Margin [-393, +393]}
	\label{tab_equiv}
	\centering
	\begin{tabular}{|l|c|}
		\hline
		Difference in Mean & -472.55\\
		\hline
		\multicolumn{2}{|l|}{95\% Confidence Interval}\\
		\hline
		Lower Limit & -489.57\\
		\hline
		Upper Limit & -455.53\\
		\hline
		Falls in Margin of Similarity & No\\
		\hline
		Claim & Not Equivalent\\
		\hline
	\end{tabular}
\end{table}

\subsection{Experiment 2}\label{sec_exp2}

\begin{figure}[!t]
\centering
\includegraphics[width=0.48\textwidth]{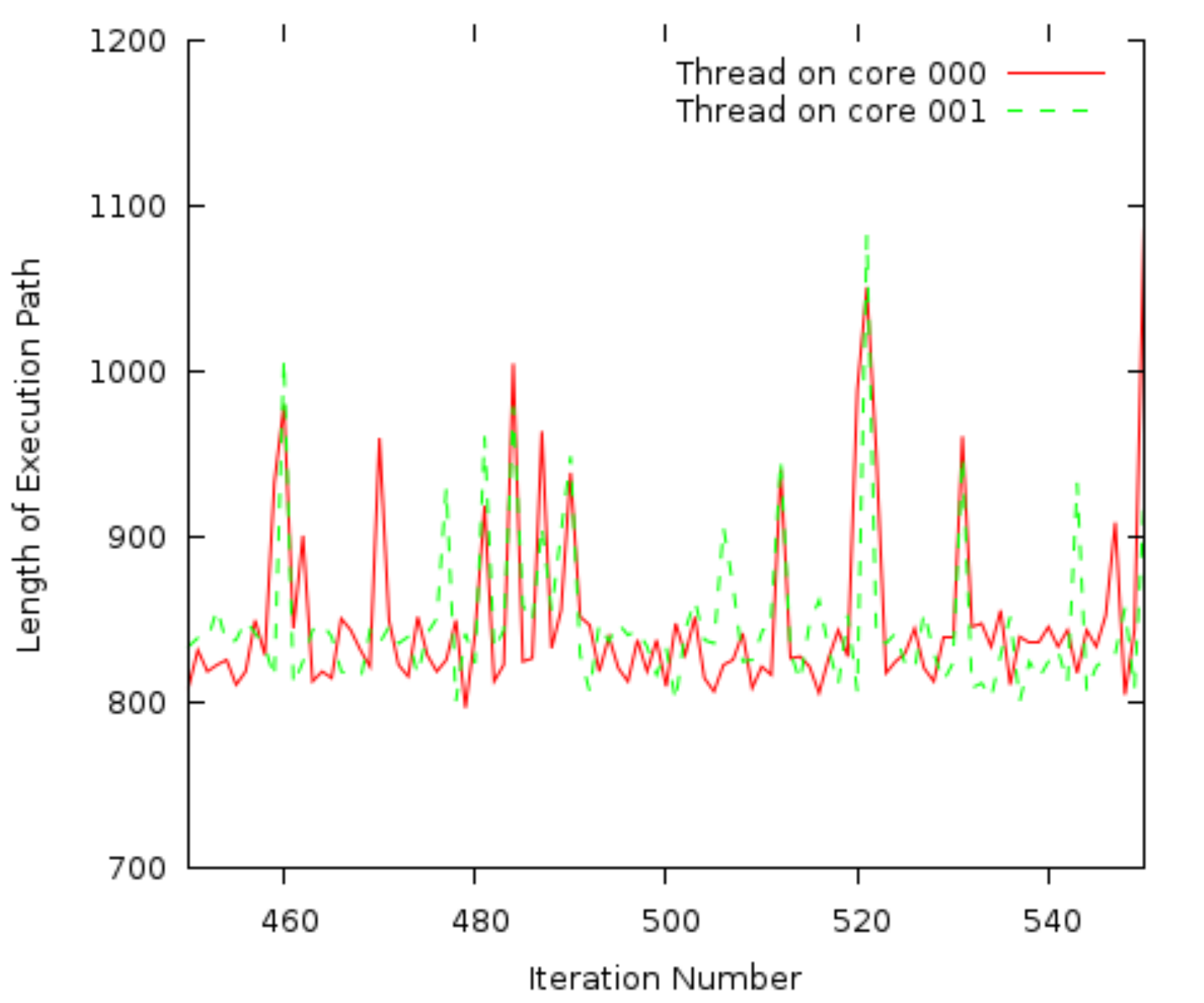}
\caption{Length of Execution Paths of Two Thread Replicas}
\label{fig_dual}
\end{figure}

We plotted, for the two threads, the variation in the lengths of the execution paths against the iteration count. For readability, we present in Fig.~\ref{fig_dual} part of the plot for the iterations 450 to 550 for the experimental run C. We noted that in a majority of the the execution instances, there was a difference in the number of functions called by the corresponding threads. For example, in the sample execution run presented, there were 5 out of the 1000 execution instances in which the corresponding instances of the two replicas had same path lengths. The absolute difference between the number of functions called by corresponding execution instances ranged from 0 to 259, with an average of 130.

\section{Discussion}\label{sec_discuss}

The experiments carried out demonstrate that there are variations in the execution paths of a user space application executing in kernel context. One possible way to characterize the variability is to use the number of kernel functions called during the execution.

The results from the execution of two replicas of the same user applications in parallel show that there is observable differences in the lengths of kernel execution paths of corresponding execution instances. The difference in the lengths of the execution paths indicate that one of the threads is executing a larger set of instructions than the other in kernel context. The corresponding execution instances thus take different distinct paths in the kernel space, which could be attributed to the `different states' of the execution environments of the threads.

Each of the system calls in the call path of the user application has a minimum (ideal) set of kernel functions it executes and would be common in all execution instances. Additional functions would be executed by the system call, for example if it requires to access exhaustible resources such as caches, contend for a resource that is held by another process or deal with an asynchronous interrupt. It is these sources of nondeterminism that results in path diversity during execution. The length of a system call path, indicated by the count of the number of functions executed during an invocation of a system call, captures this path diversity.

Comparisons of execution paths is useful, not at individual system call level, but at code units corresponding to a user's application task. The length of the path of an execution instance of the user space code fragment, is the sum of path lengths of the individual system calls invoked in the code’s execution path. Used as a measure of path diversity, the length has obvious limitations. However, it provide a good first level indicator of the level of path nondeterminism in complex execution platforms.

From the observations of the path occurrence frequencies of individual system calls in our previous work~\cite{okech2014utilizing}, we claimed that an execution path instance can categorized as either being in the frequently taken path group and rare path group. Further, the residual faults in complex software would be in the statements in these rare code paths with very high probability, since any systematic faults in the frequently taken paths would have surfaced due to the fact that the instructions in the paths are frequently executed. This claim has implications on the impact of fault activation.

Let's make an assumption that there is some unknown design fault in the code path of one of the system calls invoked by the user application. The fault would be in one of the rare path since it was not uncovered during system verification. The probability of the two threads of execution taking the same faulty path is so much less than any one of them taking the erroneous path. A 2-out-of-2 architectural configuration based on inherent diversity approach describe here would be able to detect the fault if a correct adjudication strategy is in place.

The inherent diversity approach is valid only if the taking of the untested rare paths is an independent event. If this independence assumption is violated, then the faults in these paths represent common mode faults, which would require a different detection and mitigation approach.

\section{Related Work}\label{sec_related}

The work reported in this paper achieves its objective of demonstrating that there is sufficient level of nondeterminism in program execution in kernel context through the use of dynamic program analysis, and in particular kernel tracing. Execution tracing has been applied in different scenarios such as debugging, performance monitoring, and program comprehension. In~\cite{vicente2012improving}, kernel tracing is used with the aim of understanding operating system behavior. The SIL4Linux project~\cite{wang2009sil4linux} uses kernel tracing approach to determine the variability of system calls implemented in the Linux kernel conforming to the POSIX standard over different versions. Our work, on the other hands uses the same techniques and approach to investigate the nondeterminism in program execution in kernel space.

With respect to variability of systems,~\cite{nogueira2014experimental} reports on an experimental study on the variation of execution times of program particularly the inﬂuence of operating system jitter on its variability, while~\cite{mazouz2010measuring} performed a study on the variability of execution time in multicore architectures. These two related work focus on execution time of user space applications incontrast to our work on kernel execution path.

Related to our work on the basis of using non-deterministic property of systems for protection against faults, is the INDEXYS project~\cite{eckel2010indexys}, whose stated objective is to investigate how intrinsic diversity of complex operating systems helps in detecting faults in computing platforms. They project proposers state their intension of employing architectural protection schemes to mask and/or detect random faults through temporal relaxation. We are, however, not aware of the current state of the project.

\section{Conclusion}\label{sec_concl}

We carried out two experiments to analyze the variability of the execution paths of a simple program in kernel space. In the first of the experiments we compared the paths at
the kernel level of a code fragment when the fragment was repeatedly executed in a loop. In the second experiment, two replicas of the same program were pinned to different cores of a dual core processor allowing us to compare the same code fragment when the programs are executed in parallel. The results achieved show that there is non-determinism in the execution paths at the kernel level for fault-free program execution. Further, we can say that the path taken during an instance of execution in kernel context is not fully determined by the application’s input.

The non-determinism in execution path can be attributed to the complexity of the hardware/software execution platform. This execution non-determinism, as our second experiment has shown, can manifest as diversity in replicated architectures. We claim that the inherent diversity can be used as an architectural means to provide coverage of residual faults in system software such as the operating system. This is what informs the work that we intend to carry out in the future.

In our next step, we will assess the fault detection potential of inherent diversity. We intend, through fault injection campaigns, to introduce faults in the control flow paths of system calls invoked by a program and observe if two replicas of the program in a dual channel conﬁguration exhibit the same failure. We then would follow this up with an evaluation of a real world safety application in a 2-out-of-2 conﬁguration and a realistic fault load.



\bibliographystyle{IEEEtran}

\bibliography{pokech-edcc2015}

\begin{thebibliography}{10}
\providecommand{\url}[1]{#1}
\csname url@samestyle\endcsname
\providecommand{\newblock}{\relax}
\providecommand{\bibinfo}[2]{#2}
\providecommand{\BIBentrySTDinterwordspacing}{\spaceskip=0pt\relax}
\providecommand{\BIBentryALTinterwordstretchfactor}{4}
\providecommand{\BIBentryALTinterwordspacing}{\spaceskip=\fontdimen2\font plus
\BIBentryALTinterwordstretchfactor\fontdimen3\font minus
  \fontdimen4\font\relax}
\providecommand{\BIBforeignlanguage}[2]{{%
\expandafter\ifx\csname l@#1\endcsname\relax
\typeout{** WARNING: IEEEtran.bst: No hyphenation pattern has been}%
\typeout{** loaded for the language `#1'. Using the pattern for}%
\typeout{** the default language instead.}%
\else
\language=\csname l@#1\endcsname
\fi
#2}}
\providecommand{\BIBdecl}{\relax}
\BIBdecl

\bibitem{palix2014faults}
N.~Palix, G.~Thomas, S.~Saha, C.~Calves, G.~Muller, and J.~Lawall, ``Faults in
  linux 2.6,'' \emph{ACM Transactions on Computer Systems (TOCS)}, vol.~32,
  no.~2, p.~4, 2014.

\bibitem{iec201061508Ed2}
IEC, ``61508 functional safety of electrical/electronic/programmable electronic
  safety-related systems,'' \emph{International electrotechnical commission},
  2010.

\bibitem{strigini2005fault}
L.~Strigini, ``Fault tolerance against design faults,'' in \emph{Dependable
  Computing Systems: Paradigms, Performance Issues, and Applications}, H.~Diab
  and A.~Zomaya, Eds.\hskip 1em plus 0.5em minus 0.4em\relax J. Wiley \& Sons,
  2005, pp. 213--241.

\bibitem{okech2013investigating}
P.~Okech, N.~Mc~Guire, C.~Fetzer, and W.~Okelo-Odongo, ``Investigating
  execution path non-determinism in the linux kernel,'' in \emph{Proc. 14th
  Real-Time Linux Workshop, Lugano}.\hskip 1em plus 0.5em minus 0.4em\relax
  OSADL, 2013.

\bibitem{okech2014utilizing}
P.~Okech, N.~Mc~Guire, and C.~Fetzer, ``Utilizing inherent diversity in complex
  software systems,'' in \emph{Proc. of The Australian System Safety Conference
  (ASSC2014)}.\hskip 1em plus 0.5em minus 0.4em\relax Australian Computer
  Society, Inc., 2014.

\bibitem{randell1975system}
B.~Randell, ``System structure for software fault tolerance,'' \emph{IEEE
  Transactions on Software Engineering}, vol. SE-1, no.~2, pp. 220--232, June,
  1975.

\bibitem{avizienis1977implementation}
A.~Avizienis and L.~Chen, ``On the implementation of n-version programming for
  software fault tolerance during execution,'' in \emph{Proceedings of the IEEE
  International Computer Software and Applications Conference}, vol.~77, 1977,
  pp. 149--155.

\bibitem{baudry2014multiple}
B.~Baudry and M.~Monperrus, ``The multiple facets of software diversity: Recent
  developments in year 2000 and beyond,'' \emph{arXiv preprint
  arXiv:1409.7324}, 2014.

\bibitem{rostedt2009world}
S.~Rostedt, ``The world of ftrace,'' \emph{Linux Foundation Collaboration
  Summit}, Apr. 2009.

\bibitem{mazouz2010measuring}
\BIBentryALTinterwordspacing
A.~Mazouz, S.-A.-A. Touati, and D.~Barthou, ``{Measuring and analysing the
  aariations of program execution times on multicore platforms: Case Study},''
  Tech. Rep., 2010. [Online]. Available:
  \url{https://hal.inria.fr/inria-00514548}
\BIBentrySTDinterwordspacing

\bibitem{nogueira2014experimental}
P.~E. Nogueira, R.~Matias~Jr, and E.~Vicente, ``An experimental study on
  execution time variation in computer experiments,'' in \emph{Proceedings of
  the 29th Annual ACM Symposium on Applied Computing}.\hskip 1em plus 0.5em
  minus 0.4em\relax ACM, 2014, pp. 1529--1534.

\bibitem{rostedt2010trace}
S.~Rostedt, ``trace-cmd: A front-end for ftrace,'' \emph{LWN-Linux Weekly
  News-online}, 2010.

\bibitem{vicente2012improving}
E.~Vicente, G.~Dyany, R.~Matias, and M.~de~Almeida~Maia, ``Improving program
  comprehension in operating system kernels with execution trace information.''
  in \emph{SEKE}, 2012, pp. 174--179.

\bibitem{wang2009sil4linux}
L.~Wang, C.~Zhang, Z.~Wu, N.~Mc~Guire, and Q.~Zhou, ``Sil4linux: An attempt to
  explore linux satisfying sil4 in some restrictive conditions,'' in
  \emph{Proc. of the 11th Real-Time Linux Workshop Workshop, Dresden, OSADL},
  2009.

\bibitem{eckel2010indexys}
A.~Eckel, P.~Milbredt, Z.~Al-Ars, S.~Schneele, B.~Vermeulen, G.~Csert{\'a}n,
  C.~Scheerer, N.~Suri, A.~Khelil, G.~Fohler \emph{et~al.}, ``Indexys, a
  logical step beyond genesys,'' in \emph{Computer Safety, Reliability, and
  Security}.\hskip 1em plus 0.5em minus 0.4em\relax Springer, 2010, pp.
  431--451.

\end{thebibliography}




\end{document}